\documentclass[aps,nofootinbib,draft]{revtex4}
\usepackage{amsmath}
\usepackage{graphicx}

\begin{document}

\title{Comment on soft-pion emission in DVCS}
\author{Michael C. Birse}
\affiliation{Theoretical Physics Group, School of Physics and Astronomy\\
The University of Manchester, Manchester, M13 9PL, UK\\}

\begin{abstract}
The soft-pion theorem for pion production in deeply virtual Compton scattering,
derived by Guichon, Moss\'e and Vanderhaegen, is shown to be consistent with chiral 
perturbation theory. Chiral symmetry requires that the nonsinglet operators 
corresponding to spin-independent and spin-dependent parton distributions have 
the same anomalous dimensions in cases where those operators are related by chiral 
transformations. In chiral perturbation theory, their scale-dependences can thus be 
absorbed in the coefficents of the corresponding effective operators, without 
affecting their chiral structures. 
\end{abstract}
\maketitle

\vspace{10pt}

Deeply virtual Compton scattering (DVCS) promises to provide an 
important new window on the structure of the nucleon, in the form of
generalised parton distributions (GPD's) \cite{gpd}. However, as
pointed out by Guichon, Moss\'e and Vanderhaegen \cite{gmv}, the related 
process $\gamma^*+p\rightarrow \gamma+N+\pi$, where a low-energy pion is 
produced, may be hard to disentangle experimentally from DVCS and so could 
contaminate any determination of GPD's. Guichon {\em et al.}~used a soft-pion
theorem to relate the amplitude for this process to the same GPD's that 
appear in the amplitude for DVCS. This approach has been criticised by
Chen and Savage \cite{cs}, as being inconsistent with chiral perturbation
theory (ChPT). More recently, it has been defended by Kivel, Polyakov and
Stratman \cite{kps}. Here I use the evolution of the spin-dependent parton
distributions, as calculated in Refs.~\cite{mvn,vog}, to show that 
the soft-pion theorems and ChPT are consistent, and that the results
of Refs.~\cite{gmv,kps} are correct.

The relevant operators for this discussion are the twist-2 ones corresponding 
to the moments of the nonsinglet quark distributions,
\begin{eqnarray}
\theta^{(n)a}_{V,\mu_1\dots\mu_n}&=&(i)^{n-1}\bar q\gamma_{\{\mu_1}
D_{\mu_2}\cdots D_{\mu_n\}}\tau^a q\cr
\theta^{(n)a}_{A,\mu_1\dots\mu_n}&=&(i)^{n-1}\bar q\gamma_{\{\mu_1}
D_{\mu_2}\cdots D_{\mu_n\}}\gamma_5 \tau^a q,
\end{eqnarray}
where the Lorentz tensors have been symmetrised and made traceless. (For
simplicity I consider here only two flavours of quark and so these operators
are isovector.)
The structures of the corresponding matrix elements in heavy-baryon chiral 
perturbation theory (HBChPT) are
\begin{eqnarray}
\theta^{(n)a}_{V,\mu_1\dots\mu_n}&\rightarrow &M^{n-1}A^{(n)}
v_{\{\mu_1}\cdots v_{\mu_n\}}\bar N\tau^a_+N
+M^{n-1}B^{(n)}v_{\{\mu_1}\cdots v_{\mu_{n-1}}
\bar NS_{\mu_n\}}\tau^a_-N\cr
\theta^{(n)a}_{A,\mu_1\dots\mu_n}&\rightarrow&M^{n-1}C^{(n)}v_{\{\mu_1}\cdots 
v_{\mu_{n-1}}\bar NS_{\mu_n\}}\tau^a_+N
+M^{n-1}D^{(n)}v_{\{\mu_1}\cdots v_{\mu_n\}}\bar N\tau^a_-N,
\end{eqnarray}
where
\begin{equation}
\tau^a_\pm=\frac{1}{2}\left(u^\dagger\tau^a u\pm u\tau^a u^\dagger\right),
\end{equation}
$u$ is the usual square root of the matrix $U=\exp[i\tau\cdot\phi]$ of pion 
fields in ChPT, and
$S_\mu$ is the HBChPT spin operator \cite{bkm}. I have followed here the
notation of Ref.~\cite{cs}, omitting the $\Delta$ term which is irrelevant
to the present discussion.
The coefficients $A^{(n)}$ are given by the moments of the nonsinglet, 
spin-independent quark distributions, 
\begin{equation}
q^{-}(x)=u(x)-\bar u(x)-d(x)+\bar d(x).
\end{equation}
Similarly the coefficients $C^{(n)}$ are given by the moments of the
spin-dependent distributions, 
\begin{equation}
\Delta q^{+}(x)=\Delta u(x)+\Delta \bar u(x)
-\Delta d(x)-\Delta \bar d(x).
\end{equation} 
The terms involving $\tau^a_-$ involve at least one pion field. Hence they do 
not appear in the quark distributions of the nucleon, but they do contribute 
to processes in which a pion is produced.

The structures of the HBChPT operators considered in Ref.~\cite{kps} are the 
same as those in Ref.~\cite{cs}, but with the additional constraints that 
$D^{(n)}=A^{(n)}$ and $B^{(n)}=C^{(n)}$. These arise from the fact that
Kivel {\em et al.}~\cite{kps} (see also Ref.~\cite{as})
construct the twist-2 operators
\begin{equation}
\theta^{(n)a}_{R,\mu_1\dots\mu_n}=\theta^{(n)a}_{V,\mu_1\dots\mu_n}
+\theta^{(n)a}_{A,\mu_1\dots\mu_n},\qquad
\theta^{(n)a}_{L,\mu_1\dots\mu_n}=\theta^{(n)a}_{V,\mu_1\dots\mu_n}
-\theta^{(n)a}_{A,\mu_1\dots\mu_n},
\end{equation}
corresponding to distributions of right- and left-handed quarks.
The corresponding ChPT operators are constructed using \cite{kps}
\begin{equation}
\tau^a_++\tau^a_-=u^\dagger\tau^au,\qquad
\tau^a_+-\tau^a_-=u\tau^au^\dagger,
\end{equation}
which transform under SU(2)$_R\times$SU(2)$_L$ according to
\begin{equation}
u^\dagger\tau^a u\rightarrow K(L,R,U)u^\dagger R^\dagger\tau^a
RuK(L,R,U)^\dagger,\qquad
u\tau^au^\dagger\rightarrow K(L,R,U)u L^\dagger\tau^a Lu^\dagger
K(L,R,U)^\dagger.
\end{equation}

Contrary to the claims of Chen and Savage \cite{cs}, this is a consistent
implementation of the constrants of chiral symmetry on the operators appearing
in the low-energy effective theory. The anomalous dimensions of the operators
$\theta^{(n)a}_{V,\mu_1\dots\mu_n}$ and $\theta^{(n)a}_{AS,\mu_1\dots\mu_n}$
are the same or, equivalently, the parton distributions $q^{-}(x)$ and 
$\Delta q^{+}(x)$ evolve according to the same splitting functions. These
results have been proved to second order in perturbative QCD \cite{mvn,vog}, 
but they are really just consequences of the chiral symmetry of the massless
quarks used in calculating the QCD evolution. (Note that, as discussed by Vogelsang 
\cite{vog}, some care is needed to show this because the representation of 
$\gamma_5$ used in dimensional regularisation does not anticommute with all 
the other $\gamma$ matrices.) These right- and left-handed nonsinglet quark 
distributions thus have well-defined QCD evolution properties.
In the low-energy effective theory, their dependences on the scale $Q^2$ can 
thus be absorbed in the coefficents of the corresponding effective operators, 
without affecting the chiral structure of those operators.

It is worth adding that there are no similar constraints on the 
corresponding singlet twist-2 operators, since they are invariant under 
SU(2)$_R\times$SU(2)$_L$. Indeed the spin-independent and spin-dependent 
singlet operators evolve quite differently in perturbative QCD, due to 
their different mixings with gluonic operators \cite{mvn,vog}.

Finally, I would emphasise the point made by Kivel {\em et al.}~\cite{kps}
that, at least for tree-level amplitudes, ChPT should automatically 
incorporate the old soft-pion theorems (such as Goldberger-Treiman,
Weinberg-Tomozawa, Kroll-Ruderman etc.). Where the older soft-pion methods
break down are cases for which pion-loop contributions are needed, most 
famously for $\pi^0$ photoproduction at threshold \cite{em}. 
A minor technical complication is that soft-pion results are normally 
derived using the divergence of the axial current
as an interpolating pion field. This differs at order $m_\pi^2$ from the pion fields 
in the commonly-used representations of the ChPT Lagrangians. Hence, for example, 
the soft-pion limit of pion-nucleon scattering in ChPT does not reproduce 
the pion-nucleon sigma commutator (which is also of order $m_\pi^2$). However,
in the present context of tree-level amplitudes at order $m_\pi^0$, neither
of these issues arises and so the results derived using soft-pion methods
\cite{gmv,kps} should be embodied in the effective Lagrangian of ChPT.

This work was carried out under the auspices of the EU Integrated Infrastructure 
Initiative ``HadronPhysics", contract no.~RII3-CT-2004-506078.

\end{document}